\documentstyle[epsf,12pt]{article}
\makeatletter
\long\def\@makecaption#1#2{%
  \vskip\abovecaptionskip
  \sbox\@tempboxa{\footnotesize #1: #2}%
  \ifdim \wd\@tempboxa >\hsize
    \footnotesize #1: #2\par
  \else
    \global \@minipagefalse
    \hbox to\hsize{\hfil\box\@tempboxa\hfil}%
  \fi
  \vskip\belowcaptionskip}
\makeatother
\begin{document}
\title{%
\vspace{-60pt}
\rightline{\Large\sf KUNS1529}
\vspace{12pt}
Exotic Shapes in ${}^{32}$S suggested by
the Symmetry-Unrestricted\\ Cranked Hartree-Fock Calculations
\thanks{
Talk presented by K.M. at the 
{\it Nuclear Structure '98}~~International Conference,
August 10-15, 1998, Gatlinburg, Tennessee.}
}

\author{Masayuki Yamagami and Kenichi Matsuyanagi\\
\it Department of Physics, Graduate School of Science,\\
Kyoto University,
\it  Kitashirakawa, Kyoto 606-8502, Japan}
\date{}
\maketitle

\begin{abstract}
\vspace*{5pt}
High-spin structure of ${}^{32}$S is investigated by means of the
cranked Skyrme-Hartree-Fock method in the three-dimensional Cartesian-mesh
representation. Some interesting suggestions are obtained:
1)~An internal structure change (toward
hyperdeformation) may occur at $I> 20$ in the superdeformed band,~~
2)~A non-axial $Y_{31}$ deformed band may appear in the yrast
line with $5\leq I \leq 13$.

\end{abstract}
\vspace{2cm}
\subsection*{Introduction}

Since the discovery of the superdeformed(SD) band in
${}^{152}$Dy, about two hundreds SD bands have been found in various 
mass (A=60, 80, 130, 150, 190) regions\cite{dobaczewski98}.
Yet, the doubly magic SD band
in ${}^{32}$S, which has been expected quite a long time
\protect\cite{sheline72,leander75}
remains unexplored, and will become a great challenge in the coming
years.

Quite recently, we have constructed a new computer code for the cranked
Skyrme Hartree-Fock (HF) calculation based on the three-dimensional
(3D) Cartesian-mesh representation, which
provides a powerful tool for exploring exotic shapes (breaking both
axial and reflection symmetries in the intrinsic states) at high spin
in unstable nuclei as well as in stable nuclei.
As a first application of this new code, we have investigated
high-spin structure of  $^{32}$S 
and obtained some interesting results on which we are going to
discuss below.

\newpage
\subsection*{Cranked Skyrme HF Calculation}

We solve the cranked HF equation

\begin{equation}
\delta<H-\omega_{rot}J_x>=0
\end{equation}
in the 3D Cartesian-mesh representation. 
We adopt the standard algorithm
\protect\cite{davies80,bonche85,bonche87,tajima96}
but completely remove various restrictions on spatial symmetries.
When we allow for the simultaneous breaking of both reflection and
axial symmetries, it is crucial to
fulfill the center-of-mass condition
\begin{equation}
<\sum_{i=1}^A x_i>=<\sum_{i=1}^A y_i>=<\sum_{i=1}^A z_i>=0,
\end{equation}
and the principal-axis condition 
\begin{equation}
<\sum_{i=1}^A x_iy_i>=<\sum_{i=1}^A y_iz_i>=<\sum_{i=1}^A z_ix_i>=0.
\end{equation}
Special care is taken to accurately fulfill the above conditions
during the iteration procedure. We solve these equations inside the
sphere with radius $R$=8[fm] and mesh size $h$=1[fm], starting 
with various initial configurations.
We use the Skyrme III interaction which has been successful in
describing systematically the ground-state quadrupole deformations
in a wide area of nuclear chart\cite{tajima96}.
Results of the calculation are presented in Figs. 1-3.
Figure 1 shows the structure of the yrast line.
The expected superdeformed(SD) band becomes the yrast for $I \geq 14$.
In addition to the SD band, we obtained an interesting band
possessing the $Y_{31}$ deformation, which appears in the yrast line
with $5\leq I \leq 13$. Let us call this band ``$Y_{31}$ band.''
The calculated angular momentum $I$ and deformation $\delta$
for the SD band and the $Y_{31}$ band are shown in Figs. 2 and 3
as functions of the rotational frequency $\omega_{rot}$.
Below we shall first discuss the SD band and then about the $Y_{31}$ band. 

\begin{figure}[t]
\epsfxsize=15cm
\centerline{\epsffile{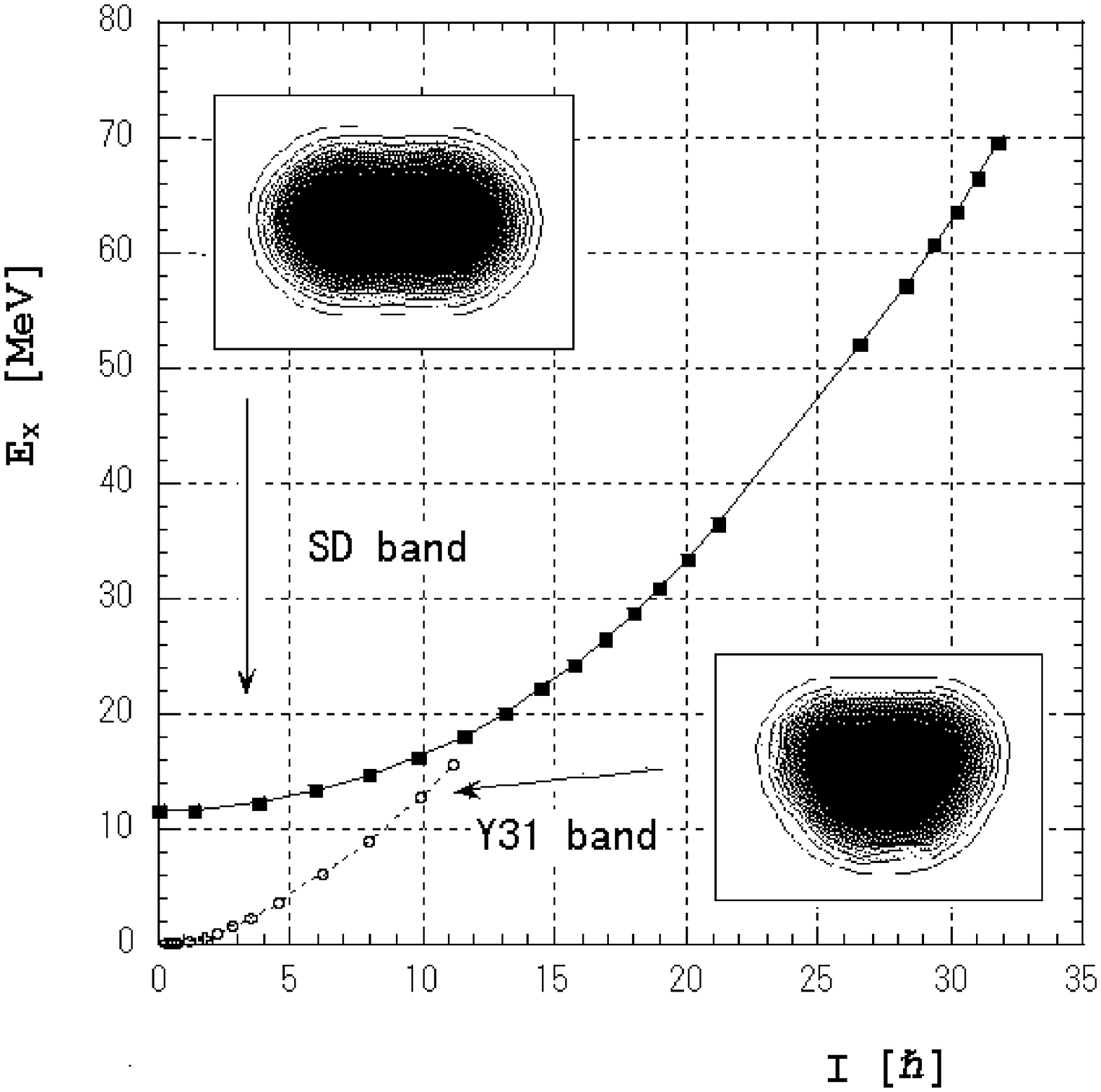}}
\caption{
Excitation energy versus angular-momentum plot of the yrast structure 
of $^{32}$S calculated with the Skyrme III interaction.
Density distributions projected on the plane perpendicular to the 
rotation axis are shown, as insets, for 
the SD band (solid line with filled squares). 
and the $Y_{31}$ band (broken line with open circles).
}
\vspace*{10pt}
\label{fig1}
\end{figure}

\begin{figure}[t]
\epsfxsize=15cm
\centerline{\epsffile{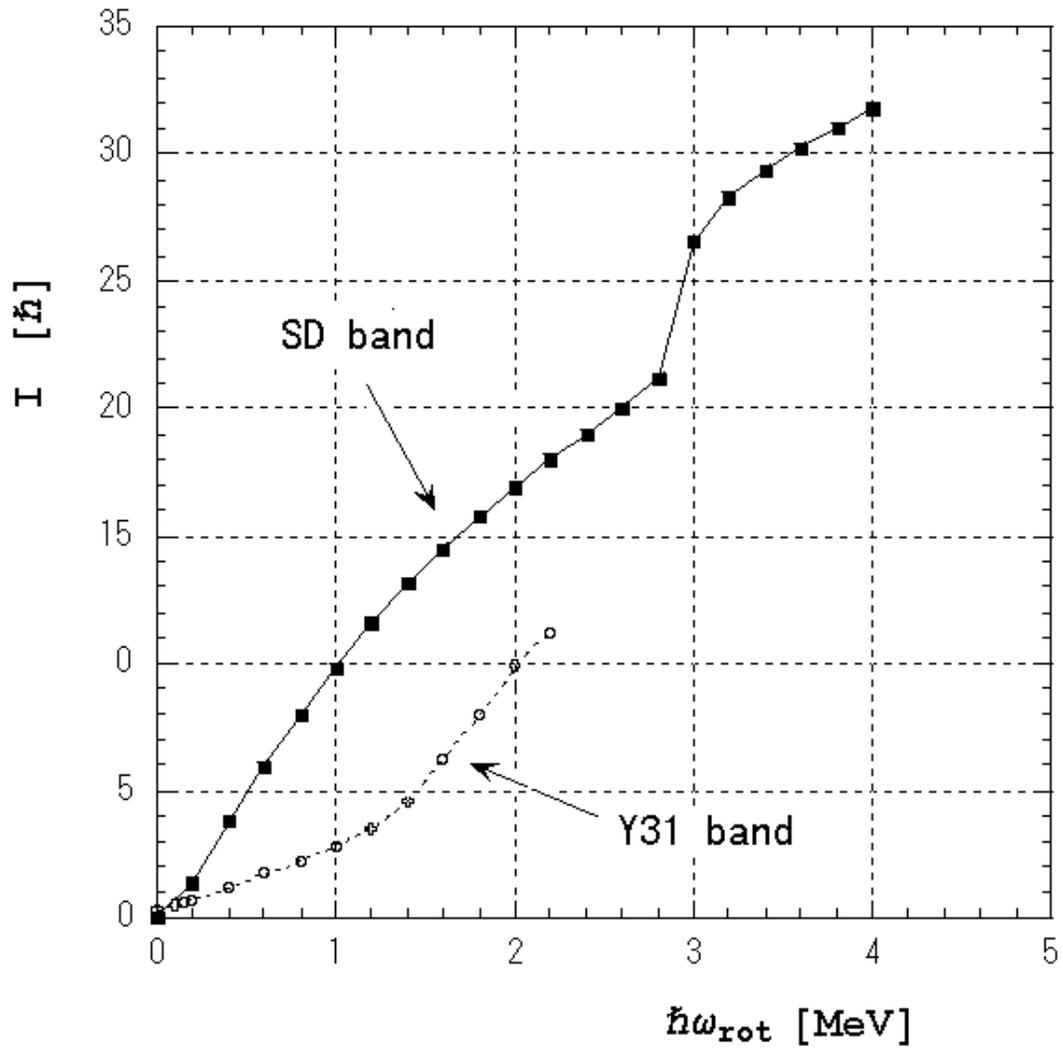}}
\caption{
Angular momenta $I$ plotted as a function of the rotation
frequency $\omega_{rot}$ 
for the SD band (solid line with filled squares) 
and the $Y_{31}$ band (broken line with open circles).
}
\vspace*{10pt}
\label{fig2}
\end{figure}

\begin{figure} [t] 
\epsfxsize=15cm
\centerline{\epsffile{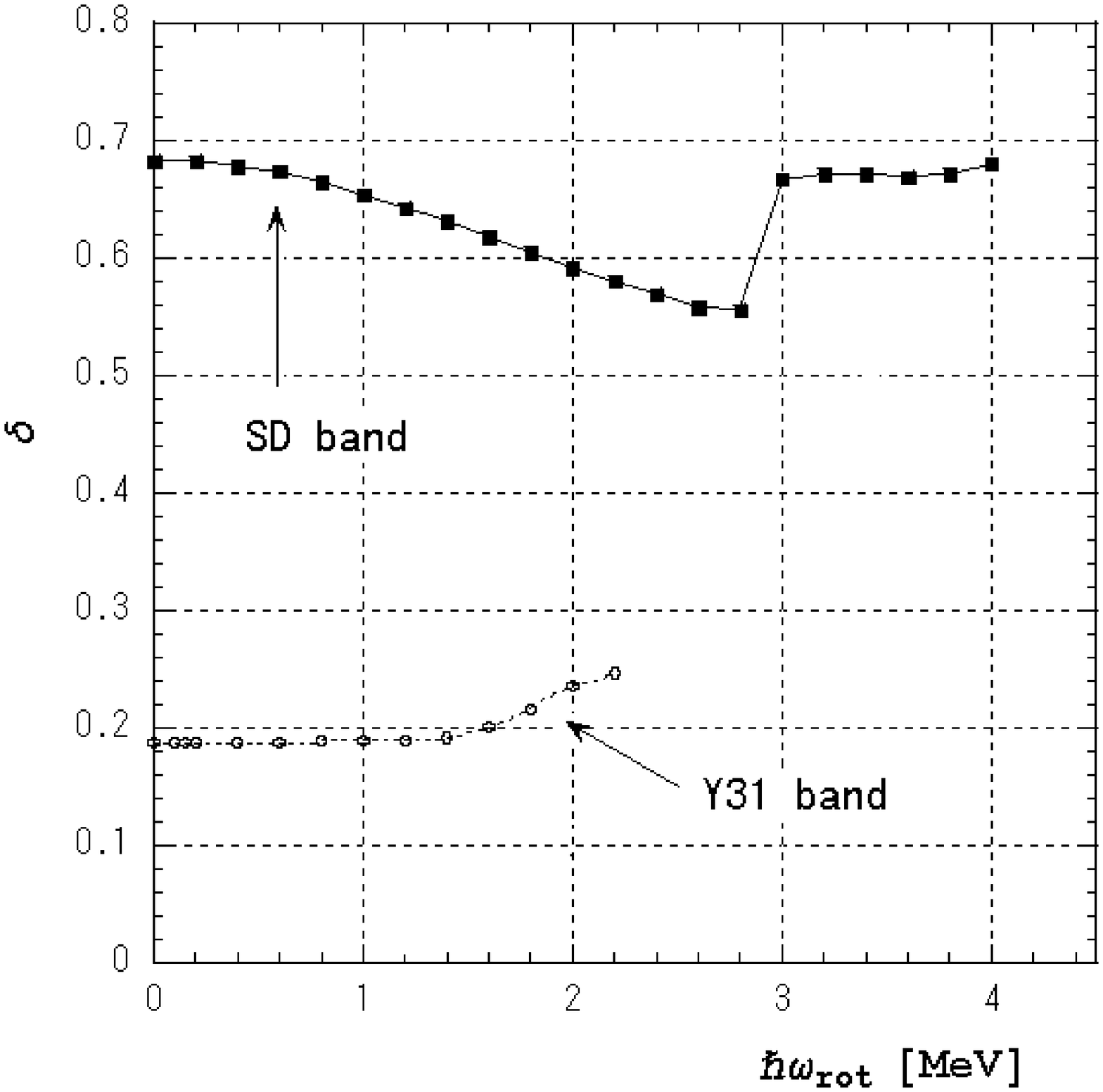}}
\caption{
Deformation $\delta$ plotted as a function of the rotation
frequency $\omega_{rot}$ for 
for the SD band (solid line with filled squares) 
and the $Y_{31}$ band (broken line with open circles).
$\delta$ is defined as
$
\delta=\frac 34~<\sum_{i=1}^A (2x_i^2-y_i^2-z_i^2)>/
<\sum_{i=1}^A (x_i^2+y_i^2+z_i^2)>
$.
}
\vspace*{10pt}
\label{fig3}
\end{figure}

\subsection*{High-Spin Limit of the Superdeformed Band}

The SD band is obtained from $I=0$ to about $I=20\hbar$.
The potential energy surface for the SD state at $I=0$ is
shown in Fig. 4. We see that the excitation energy of the SD state
at $I=0$ is about 12 MeV. 
\footnote{The rotational zero-point energy corrections are evaluated
to be -4.3 MeV and -1.1 MeV for the SD and the ground-state
configurations, respectively.
If we take these corrections into account, the excitation energy
becomes about 9 MeV.}
It becomes the yrast above $I=14\hbar$.

A particularly interesting point is the behavior of the SD band
in the high-spin limit: It is clearly seen in Figs. 2 and 3 that 
a jump occurs both in the angular momentum $I$ and the deformation $\delta$
at $\omega_{rot}\simeq 3$ MeV/$\hbar$. At this point,
$I$ jumps from about 22 to 26$\hbar$, and $\delta$ increases from 
about 0.56 to 0.66.
This is due to the level crossing with the rotation-aligned
$[440]\frac 12$ orbit.
Thus the states above $I\simeq 24\hbar$ may be better characterized as 
the hyperdeformed configuration rather than the SD configuration.
Such a singular behavior of the SD band can be noticed also in the
previous cranked HF calculation with the BKN force\cite{flocard84},
but no explanation of its microscopic origin was given there.
Let us note that if we regard the SD configuration 
as to correspond to the $j$-$j$-coupling shell model $4p$-$12h$
configuration
$\pi[(f_{7/2})^2(sd)^{-6}]\otimes\nu[(f_{7/2})^2(sd)^{-6}]$
(relative to ${}^{40}$Ca) in the spherical limit, 
the maximum angular momentum that can be
generated by aligning the single-particle angular momenta 
toward the direction of
the rotation axis is $24\hbar$, and thus ``the SD band termination''
may be expected at this angular momentum.
Interestingly, our calculation suggests that a crossover to the
hyperdeformed band takes place just at this region of the yrast line.

\begin{figure}[t] 
\epsfxsize=15cm
\centerline{\epsffile{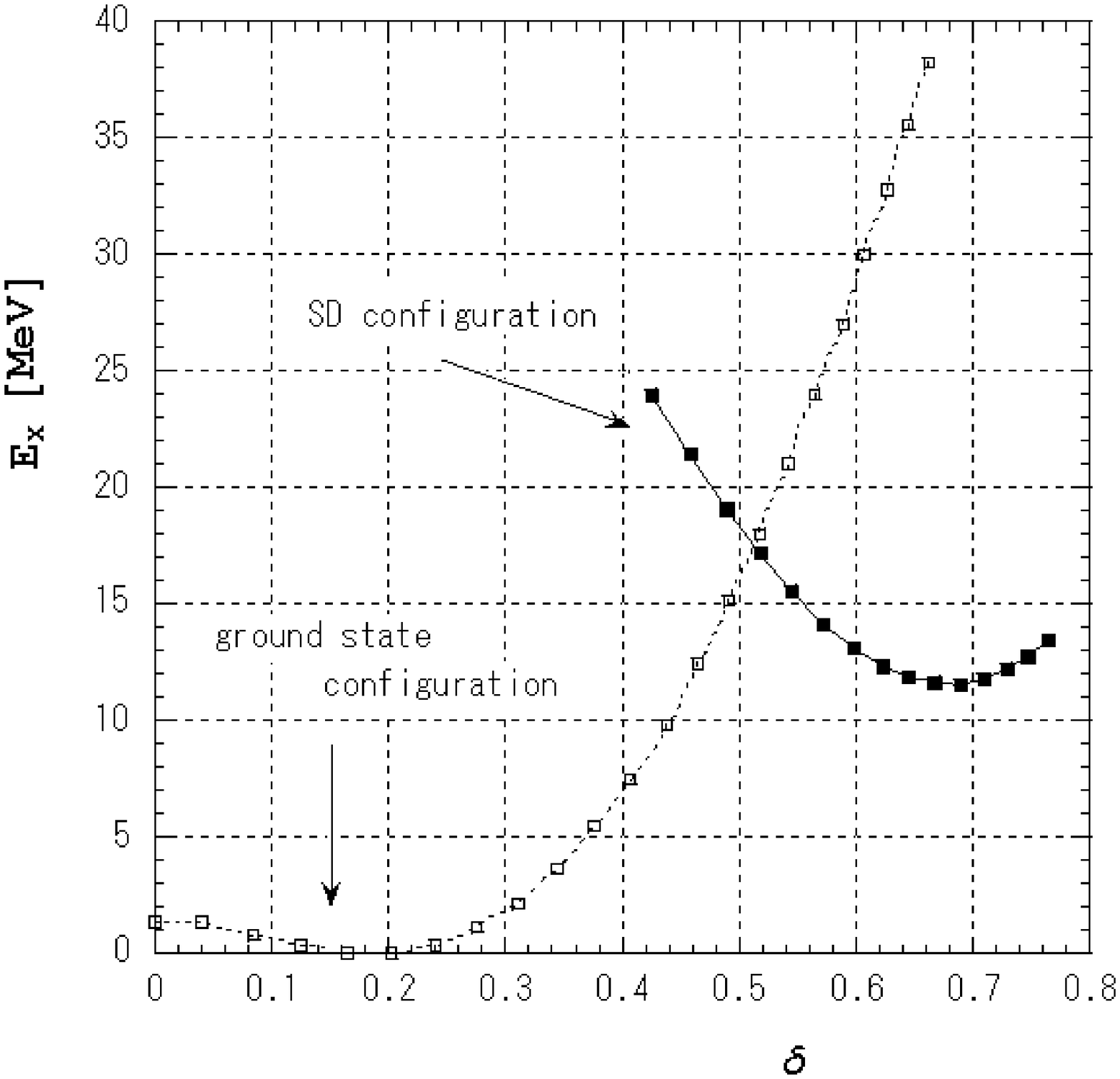}}
\caption{
Potential energy surface for the SD configuration at $I=0$
(solid line with filled squares)
relative to that for
the ground state configuration (dotted line with open squares).
This calculation was done by means of the constrained HF
procedure\protect\cite{flocard73}.
}
\vspace*{10pt}
\label{fig4}
\end{figure}

\subsection*{Effects of Time-Odd Components}

It would be interesting to examine the effect of rotation-induced
time-odd components in the mean field.
In Fig. 5 we compare the results of calculation with and without the
time-odd components. From this figure we can easily confirm that
the dynamical moment of inertia $J^{(2)}=\partial I/\partial \omega_{rot}$
of the SD band increases about 30\% due to the time-odd components.
This increase is well compared with
the effective-mass ratio $m/m^*=1/0.76 \simeq 1.3$ 
for the Skyrme III interaction, and seems to be consistent
with what expected from the restoration of the local Galilean
invariance\cite{BM}
(more generally speaking, local gauge invariance\cite{dobaczewski95})
of the Skyrme force; namely, the major effect of the time-odd
components is to restore the decrease of the moment of inertia
due to the effective mass $m^*$ and bring it back to the value for the
nucleon mass $m$.

\begin{figure} [t]
\epsfxsize=15cm
\centerline{\epsffile{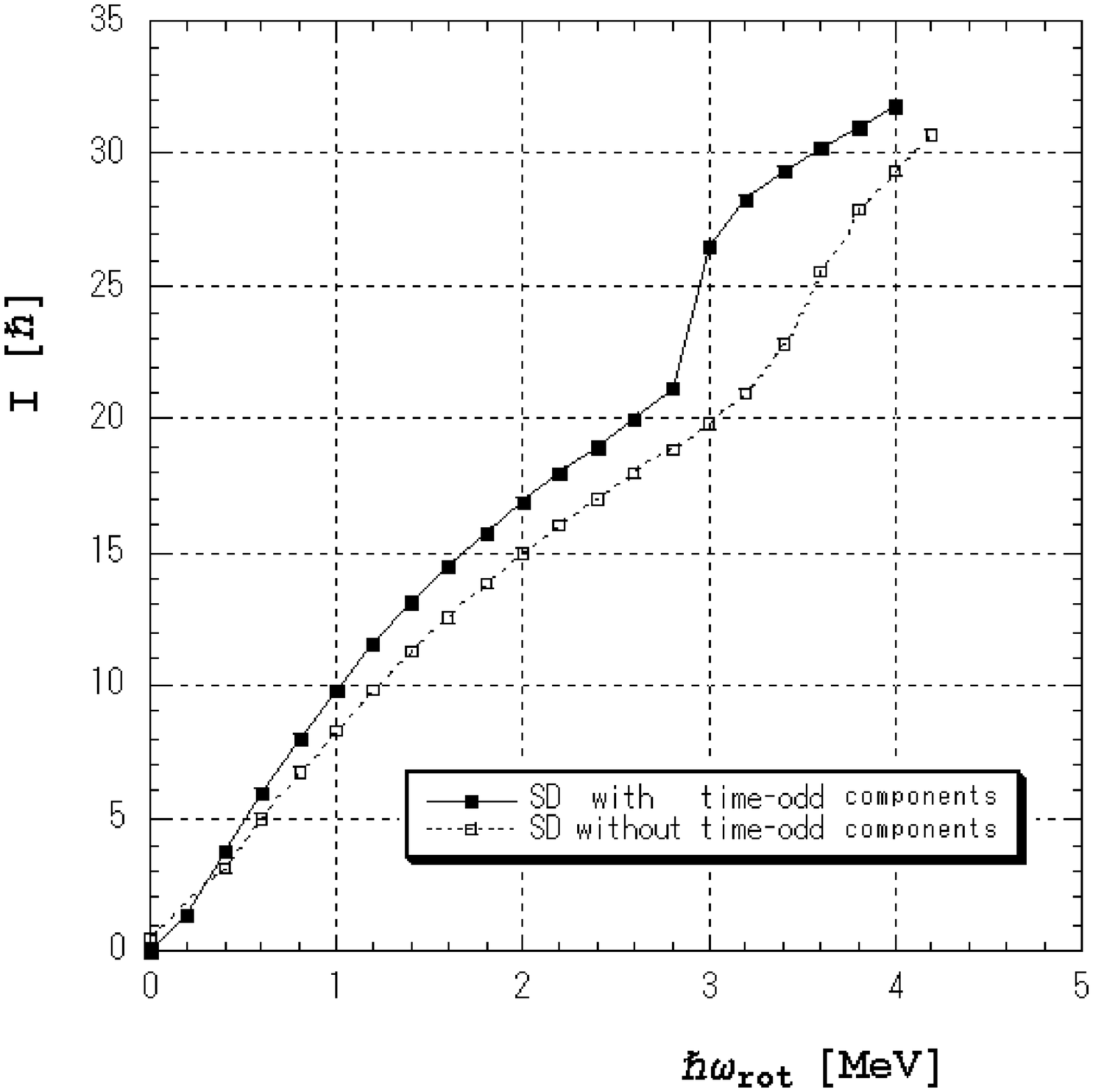}}
\caption{
Angular momenta $I$  of the SD band plotted as a function of
the rotation frequency $\omega_{rot}$. The solid line
with filled squares (dotted line with open squares)
indicates the result with(without) the time-odd components. 
}
\vspace*{10pt}
\label{fig5}
\end{figure}

\begin{figure}[t]
\epsfxsize=15cm
\centerline{\epsffile{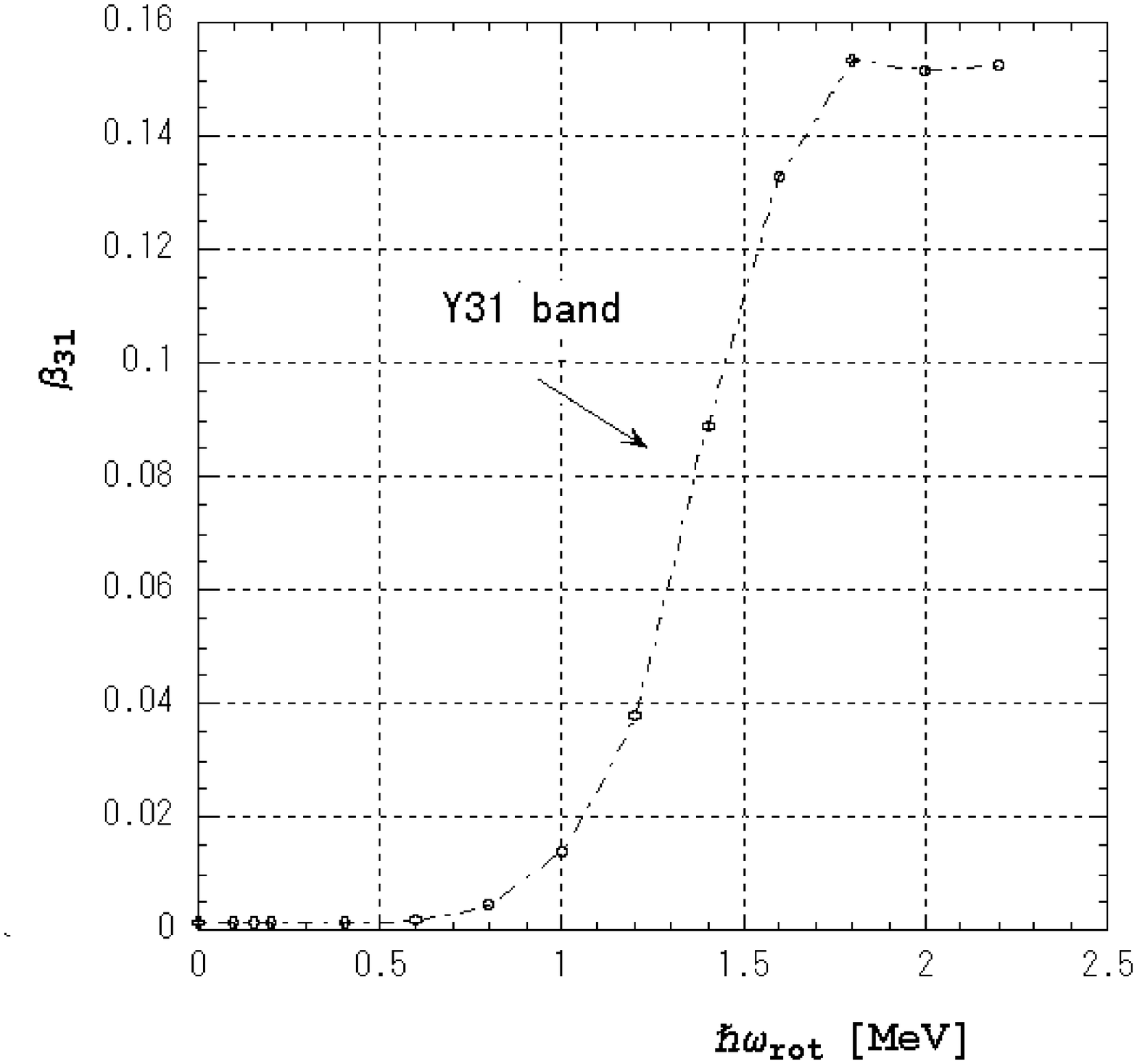}}
\caption{
Non-axial octupole deformation $\beta_{31}$ of the $Y_{31}$ band, 
plotted as a function of the rotation frequency $\omega_{rot}$. 
$\beta_{31}$ is defined through the mass-octupole moments
in the usual manner
.}
\vspace*{10pt}
\label{fig6}
\end{figure}

\subsection*{$Y_{31}$ Deformation}

As noticed in Fig. 1, we found that a non-axial
$Y_{31}$ deformed band ($\delta\simeq 0.2$ and
$\beta_{31}=0.1\sim 0.15$) appears in the yrast line with
$5\leq I \leq 13$.  
It should be emphasized that this band does not exist at $I=0$
but emerges at high spin:
As shown in Fig. 6, the $Y_{31}$ deformation quickly rises when
$\omega_{rot}$ exceeds 1 MeV/$\hbar$.

Formation mechanism of this 
band is well described as a function of angular momentum $I$ by means
of the new cranked HF code allowing for the simultaneous breaking of both 
axial and reflection symmetries.
We found that this band emerges as a result of the strong coupling between 
the rotation-aligned $[330]\frac 12$ orbit and the $[211]\frac 12$ orbit. 
The matrix element of the $Y_{31}$ operator between these
single-particle states is large, since they satisfy the selection rule 
for the asymptotic quantum numbers $(\Delta\Lambda=1, \Delta n_z=2)$.

\subsection*{Conclusions}

We have investigated high-spin structure of ${}^{32}$S by means of the
cranked Skyrme HF method in the 3D Cartesian-mesh
representation, and suggested that\\
1)~an internal structure change (toward
hyperdeformation) may occurs at $I> 20$ in the superdeformed band,\\
2)~a non-axial $Y_{31}$ deformed band may appear in the yrast
line with $5\leq I \leq 13$.

We have obtained similar results also in calculations with the Skyrme
M$^*$ interaction. More detailed study including dependence on
effective interactions is in progress.

\end{document}